# Stability of Non-Linear Integrable Accelerator*

I. Batalov#, A. Valishev, Fermi National Accelerator Laboratory, Batavia IL, 60510, U.S.A.

## Abstract

The stability of non-linear Integrable Optics Test Accelerator (IOTA) model developed in [1] was tested. The area of the stable region in transverse coordinates and the maximum attainable tune spread were found as a function of non-linear lens strength. Particle loss as a function of turn number was analyzed to determine whether a dynamic aperture limitation present in the system. The system was also tested with sextupoles included in the machine for chromaticity compensation. A method of evaluation of the beam size in the linear part of the accelerator was proposed.

## Introduction

The non-linear Integrable Optics Test Accelerator described in [1] is comprised of four periods (Figure 1). Each period consists of a linear focusing block and non-linear lens block.

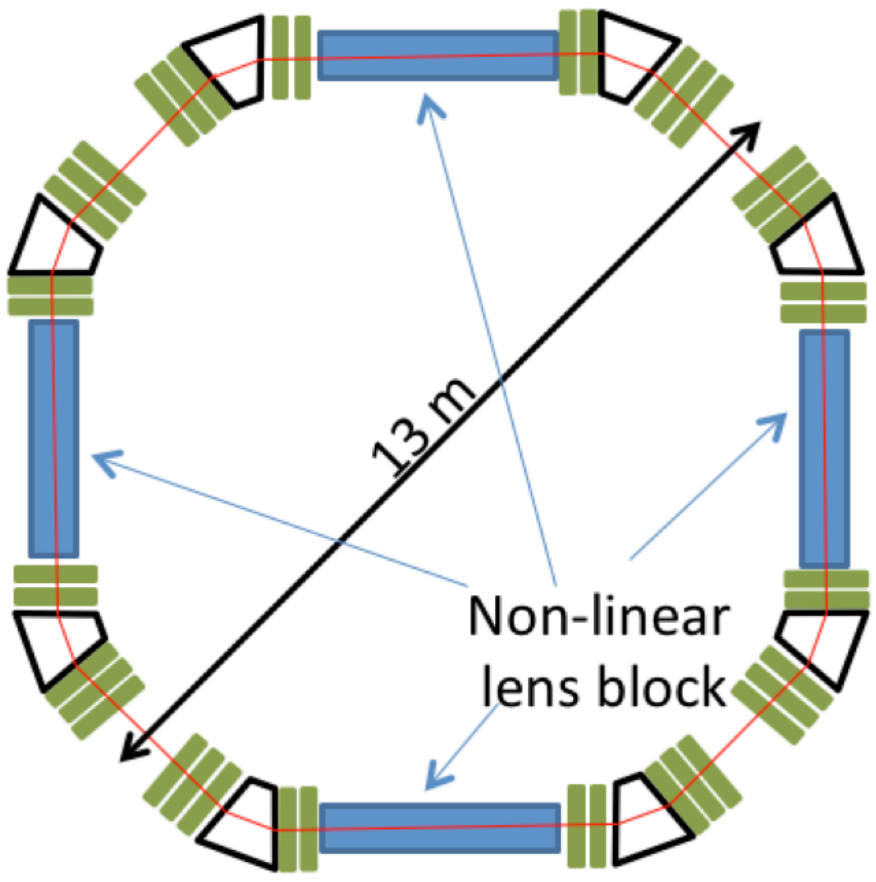

Figure 1. Layout of the non-linear Integrable Optics Test Accelerator. Quadrupoles are shown as green rectangles, bending magnets – as white trapezoids.

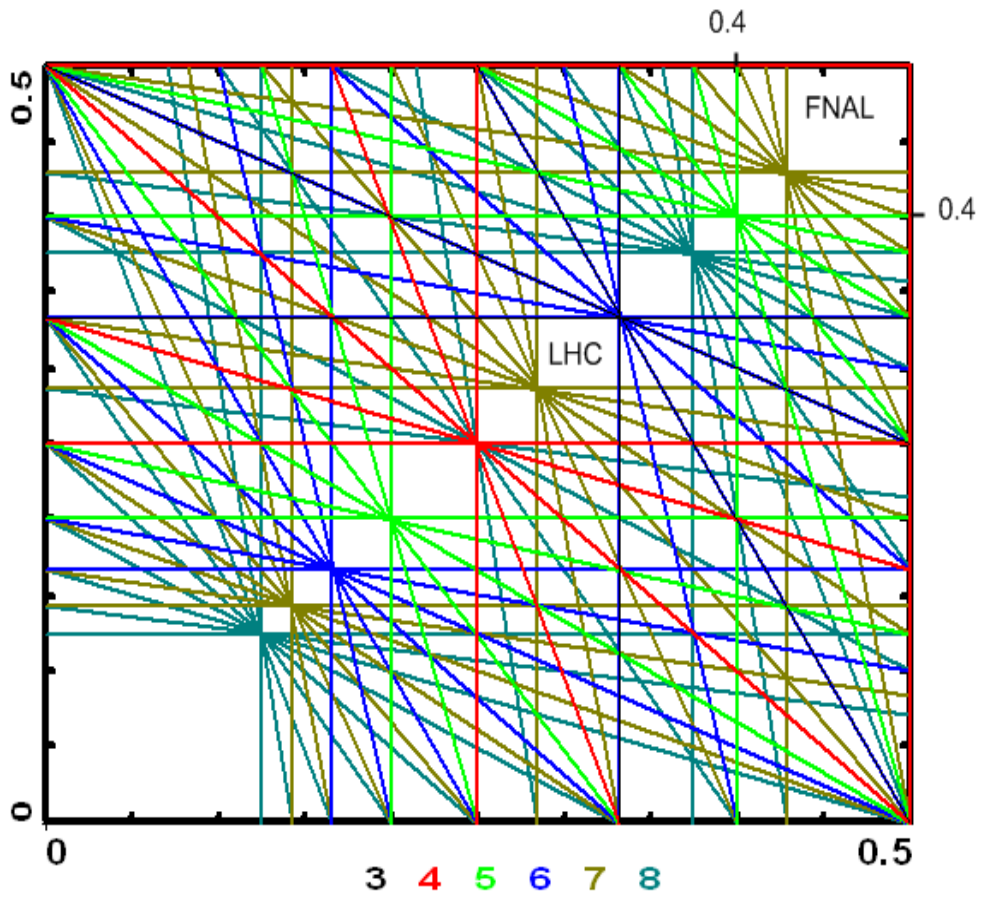

Figure 2. Map of nonlinear betatron resonances.

The advantage of this system in comparison with the conventional linear optics is the very large possible betatron tune spead and stablility against perturbations. Linear systems should operate away from resonances (Figure 2), which limits the maximum attainable tune spread. However, the wider tune spread the more stable is the beam against coherent instabilities due to Landau damping. In linear accelerators, the tune spread is generated by adding multipole magnets, which limits the dynamic aperture. In a non-linear integrable system the wide tune spread could be achieved without the excitation of resonances. Because of non-linear elements betatron tune is a function of oscillation amplitude.

The goals of this work are:

1. Calculate the stable area of transverse oscillations in the IOTA design with realistic aperture restrictions.
2. Determine whether a dynamic aperture limitation is present.
3. Calculate the betatron tune spreads that could be achieved in the system.
4. Study the effect of the chromaticity correction sextupoles on particle stability and maximum tune spread.

All simulations were carried out using a special version of MAD-X [2], in which the non-linear lens was implemented.

---

* Fermilab is operated by Fermi Research Alliance, LLC under Contract No. DE-AC02-07CH11359 with the United States Department of Energy.

# ibatalov@fnal.gov, also Moscow Institute of Physics and Technology, Russia

## Apertures. Particle motion

In Table 1 the information about apertures used in the machine is shown. Simulations show that almost all lost particles are dying on the aperture of quadrupoles. Therefore this aperture is limiting. Enlarging the quadrupole apertures may increase the stable region, but construction of such quadrupoles requires more expensive superconductive technologies.

| **Element** | **aperture shape** | **dimensions** |
|---|---|---|
| quadrupole | circle | 2 inches in diameter |
| dipole | rectangle | 2x1 inches |
| non-linear lens | ellipse | x half-axis: 0.7×c×$\beta$<br>y half-axis: 7×c×$\beta$ |

Table 1. Description of element apertures in non-linear accelerator model.

Figure 3 shows the trajectories of particles in y-y' phase space at two different values of the non-linear lens strength. In the case when tn<0.5 the phase space trajectories go around the origin of coordinates. But if tn>0.5, particle trajectories do not encircle the origin of coordinates, linearized system is unstable and the beta functions cannot be found. Thus, the amplitudes of particle oscillations can only be found by particle tracking.

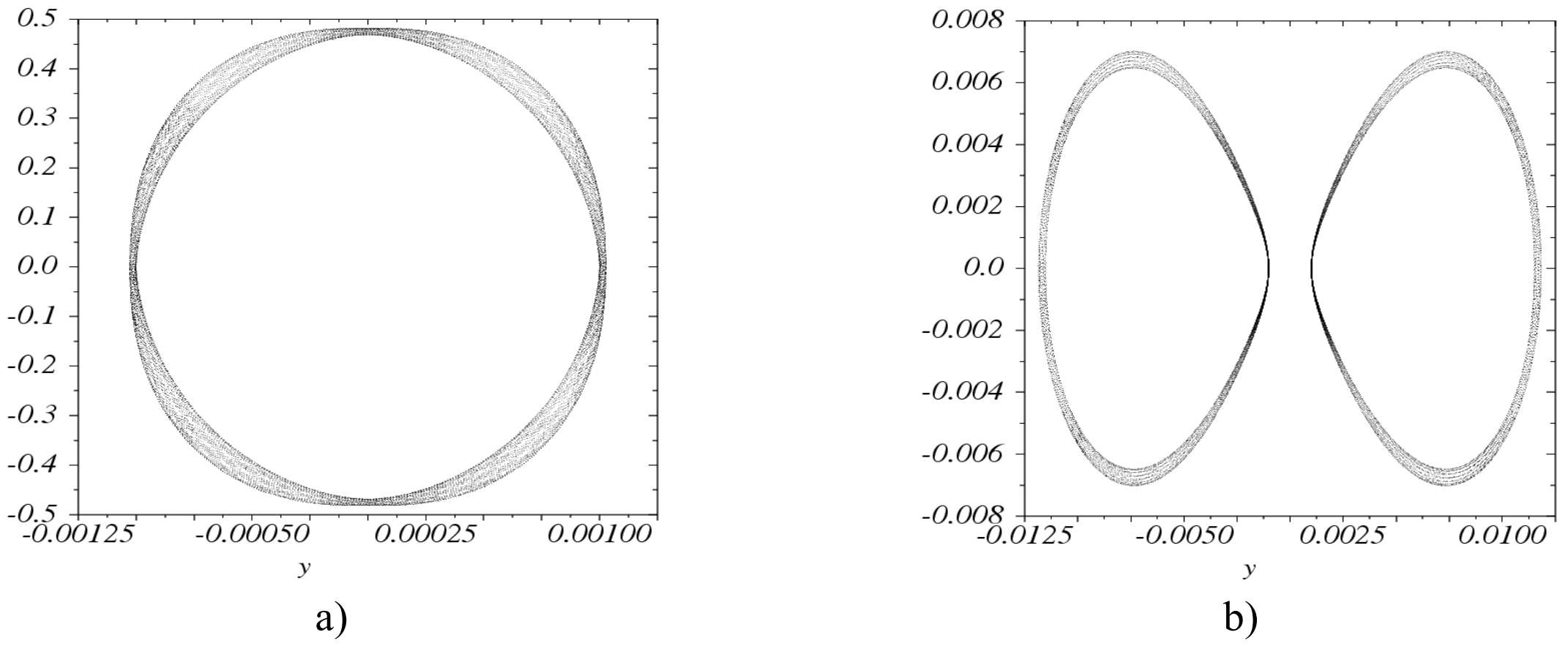

Figure 3. Particle trajectories in y-y' phase space. a) strength of non-linear lens tn = 0.4; b) tn = 1.0

## Stable area. 1-D case

The first question that arises from aperture limitations is how the size of the stable region depend on the strength of non-linear lens. To answer it all the initial coordinates of the particle except one ('x 'or 'y') were set to zero. The observation point was selected at the center of the non-linear lens block. The non-zero coordinate was varied to achieve maximum amplitude oscillation equal to the aperture. Figure 4 represents the maximum amplitude of particles' oscillations inside of non-linear lens as a function of the lens strength.

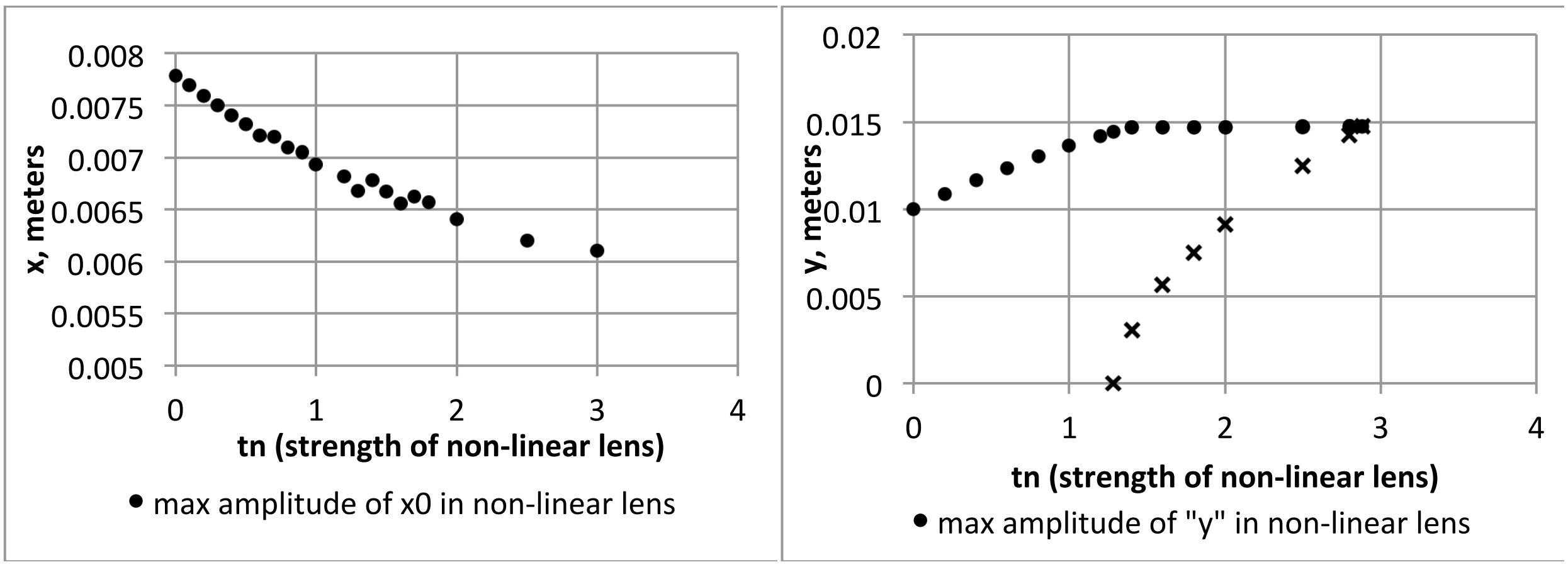

a) b)

Figure 4. a) maximum oscillation amplitude of survived particles along x-axis at the center of non-linear lens block; b) maximum and minimal amplitude along y-axis.

For the horizontal motion the maximum amplitude decreases with the increase of non-linear lens strength. For vertical motion there are 2 critical values: the maximum oscillation amplitude, which is equal to maximum initial coordinate, and minimum initial coordinate. The minimum initial coordinate can be explained with particles' trajectories in y-y' phase space (Figure 5). The non-linear potential can be represented by an expansion:

$$U = Re\left[(x+iy)^2 + \frac{2}{3} * (x+iy)^4 + \frac{8}{15} * (x+iy)^6 + \frac{16}{35} * (x+iy)^8 + \ \ldots\right]$$

The first term is the quadrupole potential, which focuses in x-direction and defocuses in y-direction. Therefore, the maximum deviation of a particle with small initial coordinate is bigger that of those with larger initial coordinate.

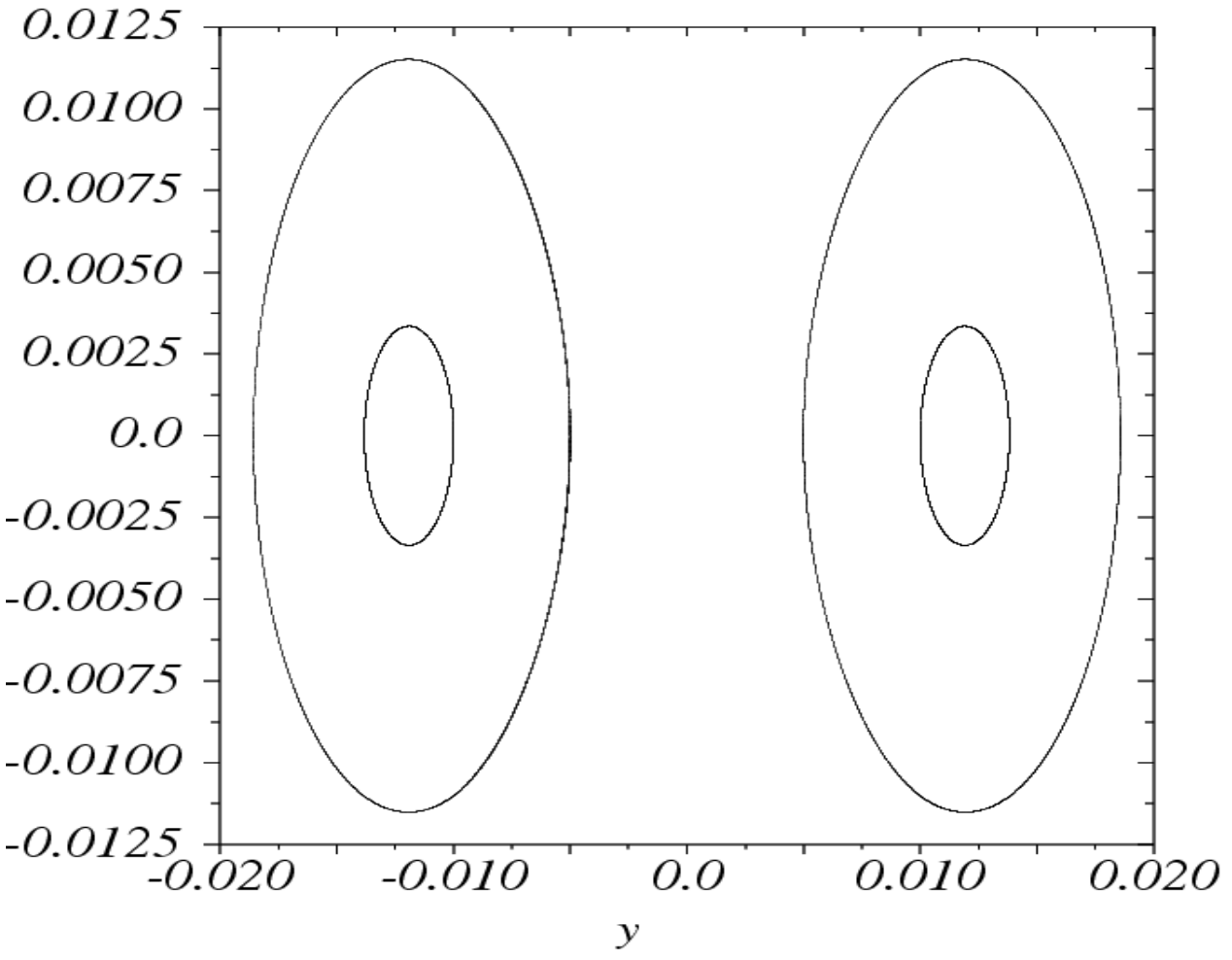

Figure 5. Trajectories of two particles at the center of non-linear lens block. Inner trajectory corresponds to higher initial y-coordinate, outer trajectory – to lower initial y-coordinate. Other coordinates are zero.

## Stable area. 2-D case

The case with only one non-zero initial coordinate does not describe the whole pool of initial coordinates. To see the real picture of stable area one should set at least 2 non-zero coordinates. Although each particle has 4 coordinates (x, x', y, y'), there is coupling between them. In Figure 5 one can see particle's trajectory in y-y' phase plane. The similar picture can be build for x-x' plane. This trajectory is the same for any particle with the initial coordinates belonging to the trajectory. Therefore, the initial point can be put anywhere on the trajectory, and we choose it in the x-y plane.

Figure 6 represents the stable areas for the particles with non-zero 'x' and 'y' initial coordinates for different strengths of the non-linear lens. One can see that the stable area splits into two parts starting from some value of strength confirming the results of 1-D case. From this figure it is also clear that there is maximum strength of non-linear lens for which particles still survive.

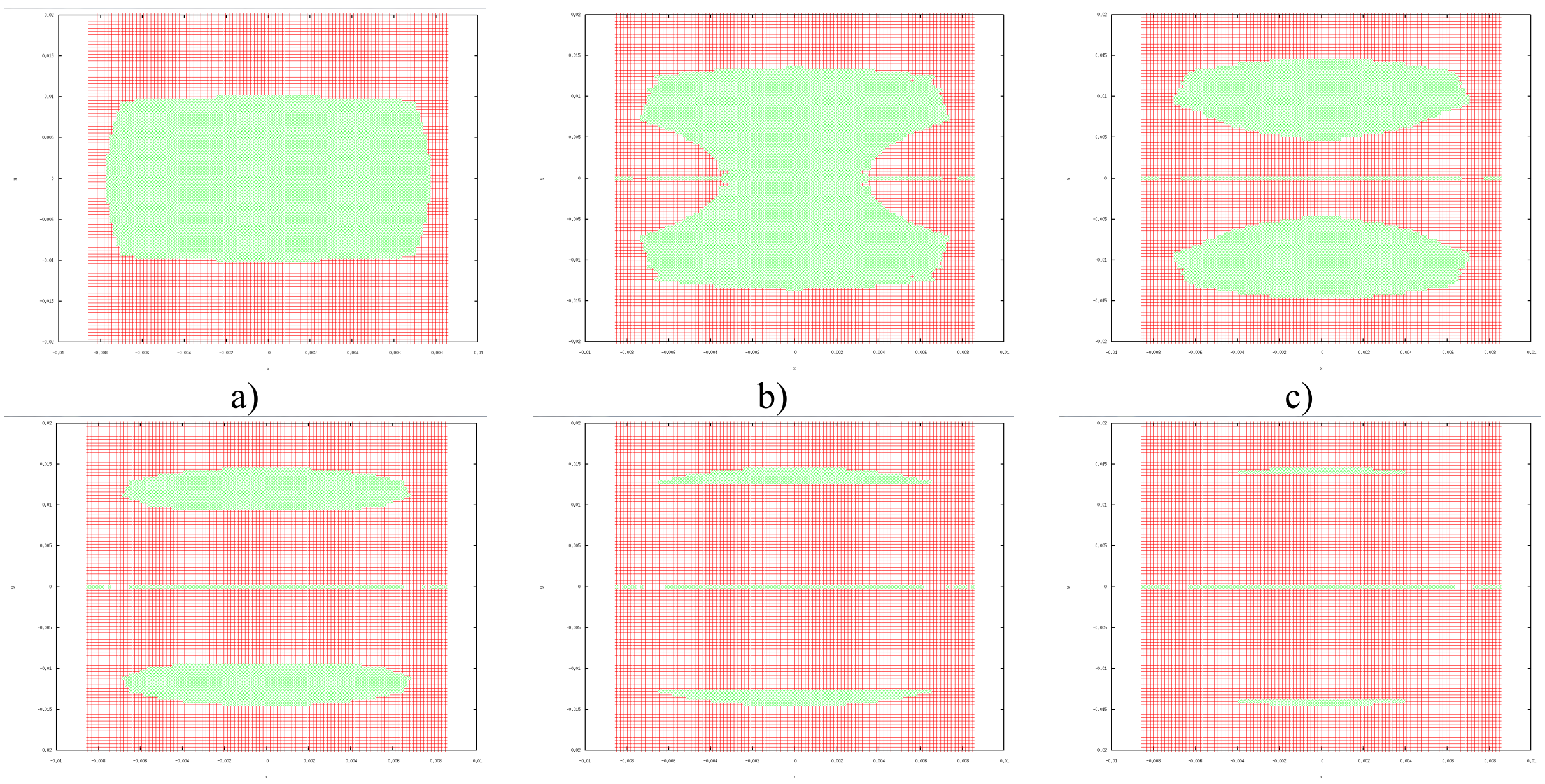

a) b) c) d) e) f)

Figure 6. x-y planes of initial coordinates of lost (red) and survived (green) particles for various non-linear lens strengths (tn): a) tn=0; b) tn=0.5 c) tn=1.0; d) tn=1.5; e) tn=2.0; f) tn=2.5

There are 2 possible reasons why particles die: mechanical and dynamic aperture. The first one is caused by geometrical restrictions of the machine. The second one is caused by non-regular chaotic motion. If the particle is lost due to mechanical aperture, it will die quickly, in 1-100 turns. If dynamic aperture is present in the system, particles will be lost even after large number of turns. Table 2 presents the number of lost particles as a function of turn number. In the case of zero non-linear lens strength there can be no dynamic aperture because the system is totally linear and ideal. Almost all particles are dying before the first 100 turns. Particle losses after larger number of turns can be explained by computational errors for the particles near the aperture and the number of such particles is comparatively small. Since for non-zero strength of the non-linear lens this number decreases, there is no dynamic aperture limitation present in the system.

| | **number of lost particles** | | | |
|---|---|---|---|---|
| **tn** | **turn 1-10** | **turn 10-100** | **turn 100-1000** | **turn 1000-2000** |
| 0 | 5768 | 76 | 8 | 4 |
| 0,5 | 5545 | 76 | 11 | 0 |
| 1 | 5610 | 132 | 6 | 2 |
| 1,5 | 6943 | 122 | 8 | 0 |
| 2 | 8437 | 24 | 8 | 4 |
| 2,5 | 9625 | 8 | 0 | 0 |
| 2,7 | 10066 | 0 | 0 | 0 |

Table 2. Number of lost particles as a function of turn interval and non-linear lens strength.

## Fourier transform. Foot prints

To describe the tune spread of the bunch in the accelerator, one moves from the physical coordinates to frequency domain by Fourier transformation. We select the main tune by finding the maximum of Fourier amplitude. Since this procedure is done for every particle in the bunch, one can plot all these maxima on the $Q_x$-$Q_y$ diagram. This picture is called the foot print and is shows the tune spread within the bunch. One of the examples of the foot print is shown in Figure 7.

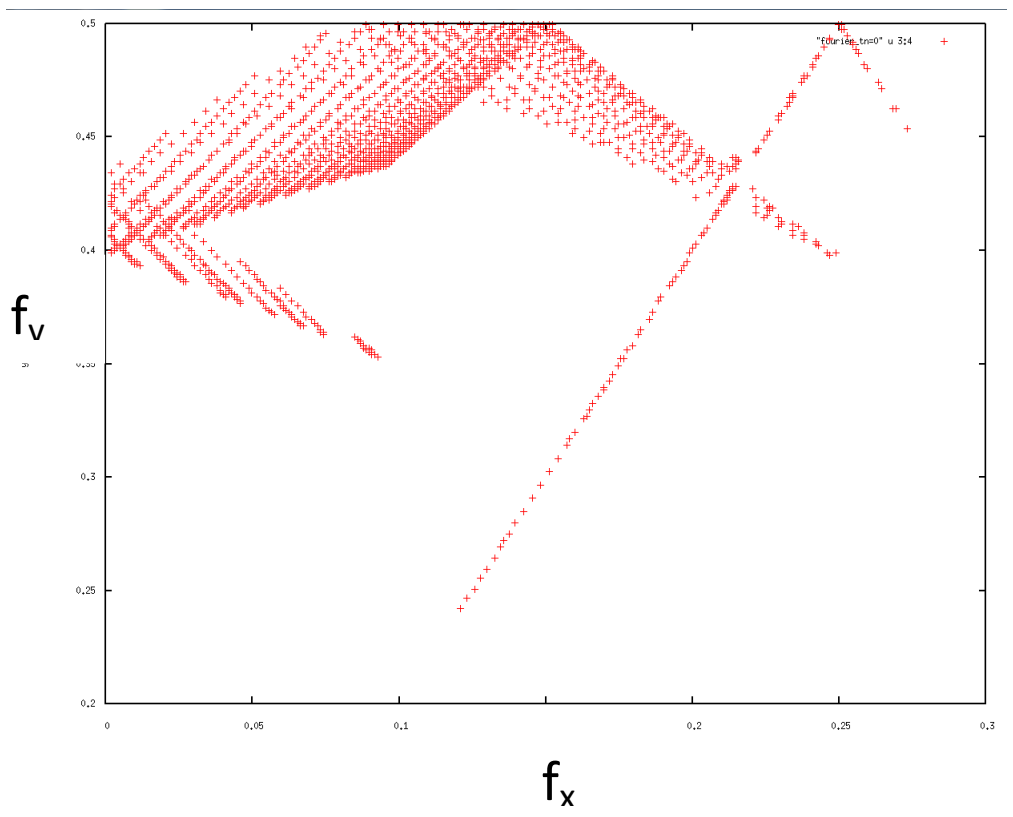

Figure 7. Foot print example for non-linear lens strength tn=1.5

Because coordinates of the particles are observed at one point of the accelerator, the discrete Fourier transform cannot show all the frequencies adequately. It appears in reflections of foot print from the boundaries of the square [0,0.5]x[0,0.5] on the plot. Reflections also mean crossing of the integer and half-integer resonances that are important for us in order to test the system stability in these most unstable regions. To measure the full tune spread one should these reflections into account. So, the real foot print shown in Figure 7 will look like a triangle.

Figure 8 presents the tune spread measured by the routine described above as a function of the non-linear lens strength. One can see that tune spreads for both x and y coordinates has maxima. The presence of these maxima is the result of interplay between the increasing tune shift and decreasing stable area as functions of non-linear lens strength. Although the tune spread eventually shrinks, the maximum value is much bigger than for linear accelerators. Therefore Landau damping should be significant and stability of the bunch to perturbations will be much stronger than in conventional linear systems.

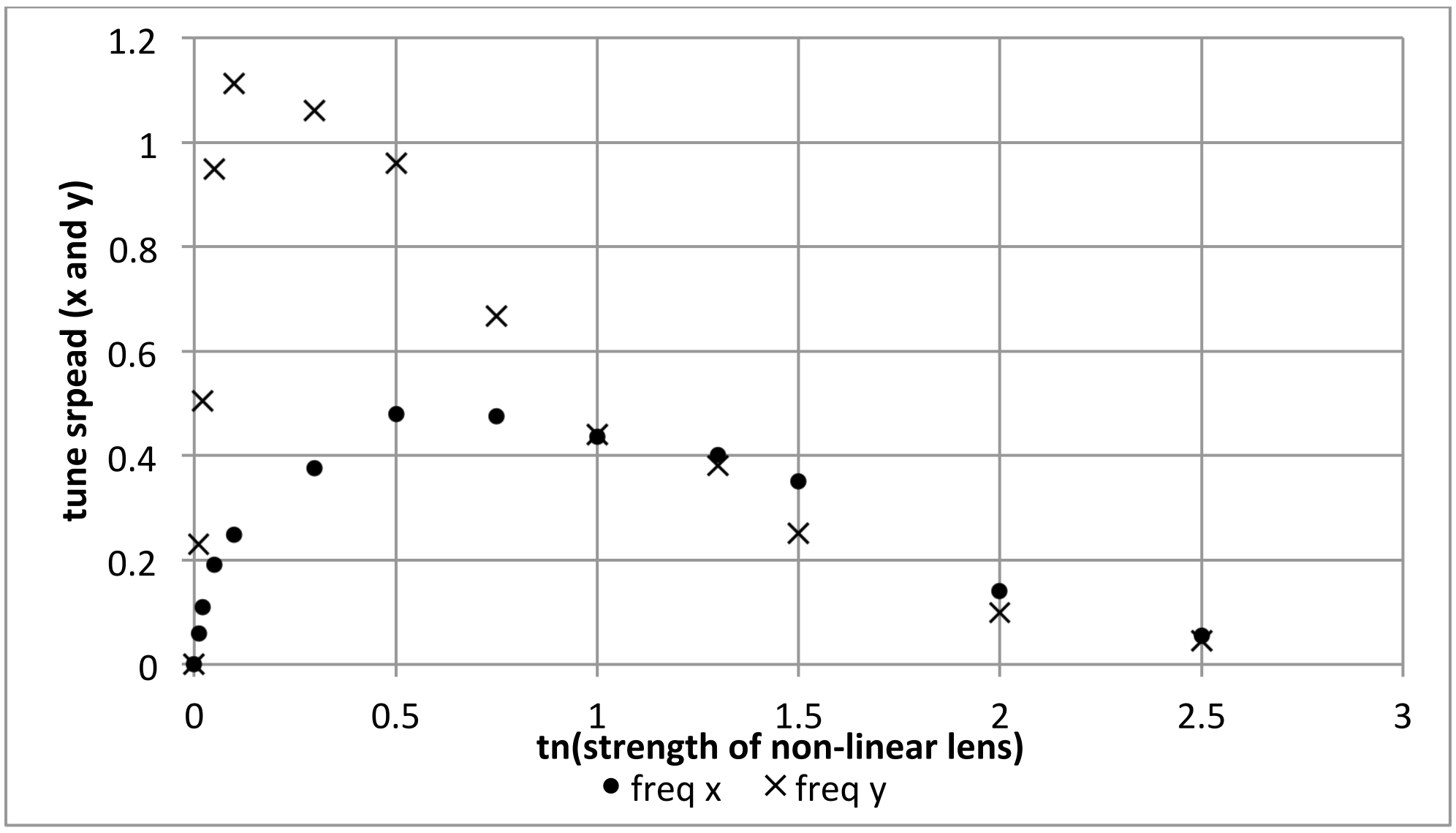

Figure 8. $Q_x$ and $Q_y$ tune spreads as functions of non-linear lens strength.

## Effect of sextupoles on tune spread

Sextupoles are used in accelerators to compensate the betatron tune chromaticity. Chromaticity appears in quadrupoles and it is negative:

$$\frac{\partial Q_x}{\partial \delta} = -\frac{1}{4\pi} \times \int_0^L \beta_x(s) k_x(s) ds$$

$$\frac{\partial Q_y}{\partial \delta} = -\frac{1}{4\pi} \times \int_0^L \beta_y(s) k_y(s) ds$$

Sextupole chromaticity is positive in one direction and negative in another :

$$\frac{\partial Q_x}{\partial \delta} = \frac{1}{4\pi} \times \int_0^L \beta_x(s) r(s) D(s) ds$$

$$\frac{\partial Q_y}{\partial \delta} = -\frac{1}{4\pi} \times \int_0^L \beta_y(s) r(s) D(s) ds$$

Consequently, adding 2 sextupole families to the system at appropriate places with correct strengths can suppress the chromaticity of quadrupoles. But sextupoles also introduce a dynamic aperture in the system. This decreases the stable area, which results in particle losses at large numbers of turns (Figure 8).

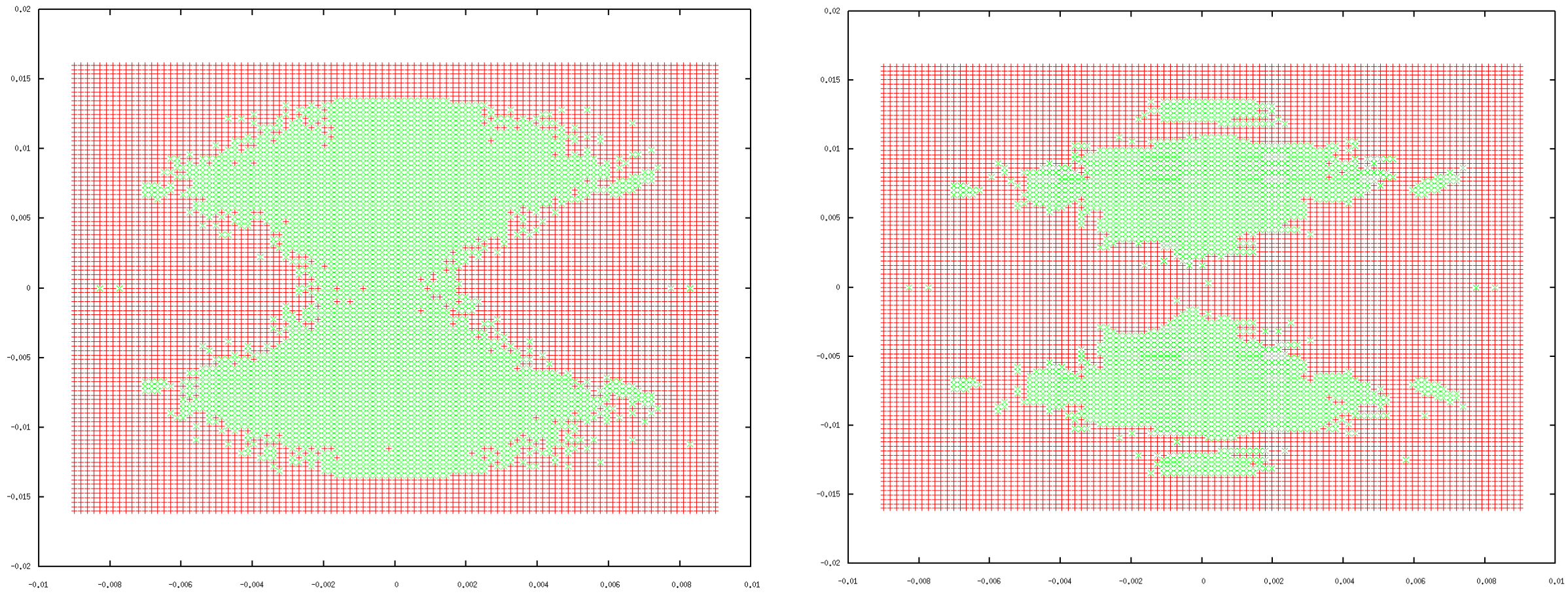

a) b)

Figure 8. x-y planes of initial coordinates with sextupoles on. Lost particles are in red and survived in green. a) 2048 turns; b) 65536 turns.

To predict the influence of chromaticity sextupoles on the system stability, the tune spread change was analyzed. For this purpose the area of initial coordinates that corresponds to survived particles was truncated in order to exclude potentially unstable particles. Figure 9 presents tune spreads in x and y directions with and without sextupoles. It is seen from the plots that tune spread in x direction decreases when sextupoles are present while the tune spread in y direction does not change significantly.

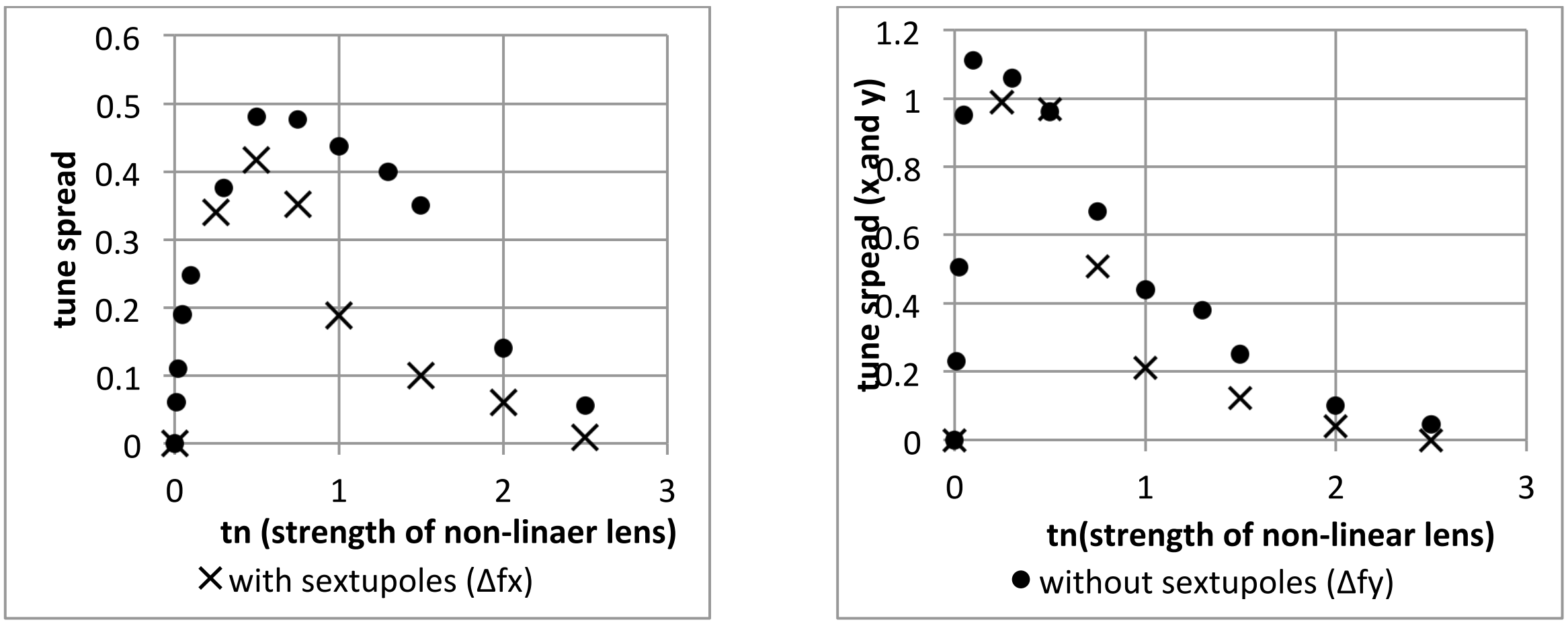

a) b)

Figure 9. Tune spreads with and without sextupoles as functions of non-linear lens strength. a) along x-axis; b) along y-axis.

## Beam size in machine arcs

Although tracking is a reliable way to predict the beam size in a non-linear accelerator, it is very slow. When the strength of non-linear lens is less than 0.5, the beam size can be evaluated with the use of beta functions. This method is much faster than tracking and allows to carry out extensive simulations (for example, adjustment of quadrupole parameters to make beam size minimal). But for the non-linear lens strength larger than 0.5, this method becomes impossible because the linearized system is unstable. However, a convenient method is needed to describe the beam at least in the linear part of the accelerator. To do this the following method is proposed. First, the beam shape at the beginning of linear section is calculated by tracking with the accelerator arc replaced with an equivalent map. Then, an ellipsoid with minimal area is circumscribed over the beam phase portrait. Finally, the alpha and beta functions are derived from the shape of the ellipsoid and used as initial values to find alpha and beta functions throughout a linear part of accelerator.

Figures 10, 11 and 12 present an implementation of this idea in 2D y-y' phase space. All x and x' coordinates in this case are equal to zero. In Figure 10 one can see two particle distributions inside the beam at the beginning of linear section. The shape shown in Fig. 10a is chosen to be more ellipse-like while the shape in Fig. 10b is chosen to be less ellipse-like. Both of these shapes were used to check the idea described above.

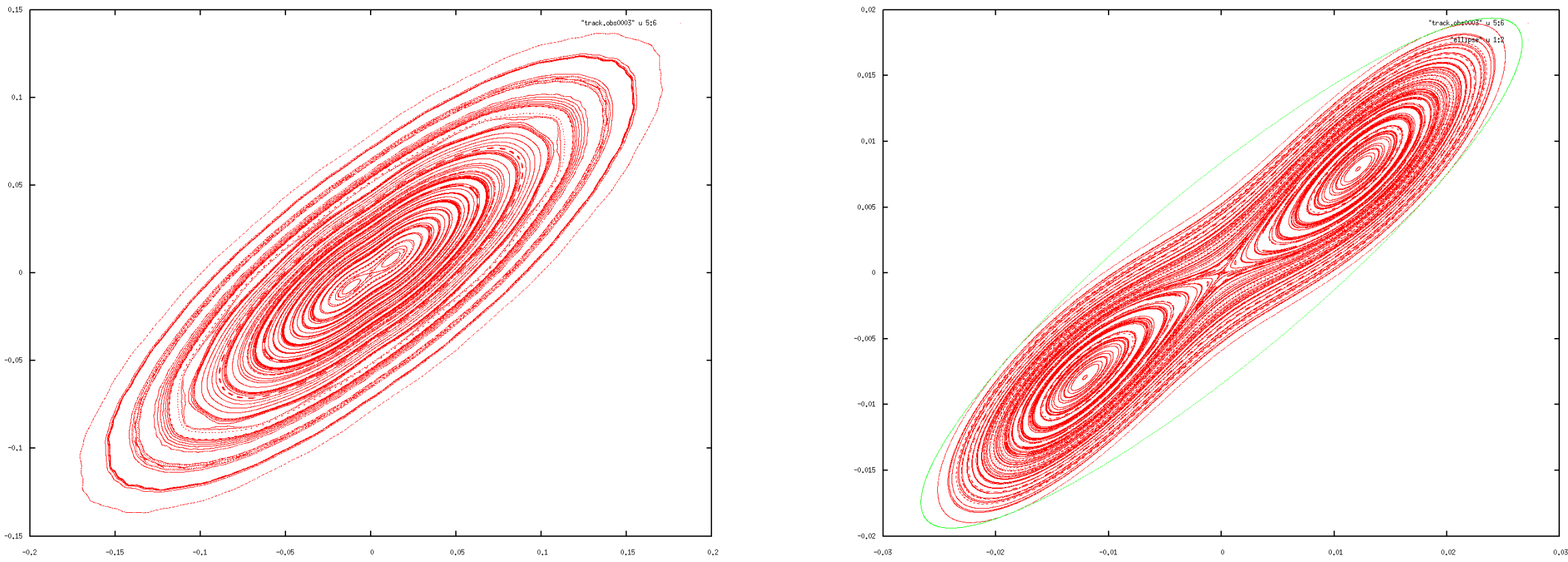

a) b)

Figure 10. Particle distribution at the beginning of linear section (y-y' plane). a) ellipse-like shape; b) not ellipse-like shape.

First, let us consider ellipse-like initial distribution. Figure 11a presents the comparison between the vertical beam size calculated by tracking and the square root of $\beta_y$ function calculated using the circumscribed ellipsoid. One can see that the plots coincide very well so the prediction of beam size given by beta-functions of circumscribed ellipsoid is accurate. Figure 11b presents squared maximum vertical beam size divided by emittance as a function of longitudinal coordinate. The 1$^{st}$ derivative of this function could be compared to 1$^{st}$ derivative of $\beta_y$ which can be found using the following expression:

$$\begin{cases}\alpha_y = -\beta_y'/2 \\ \alpha_y = -1.38\end{cases} \rightarrow \beta_y' = 2.77$$

As one can see from the Figure 11b, $(y^2_{max}/\varepsilon)' = 2.34$ which is close to the value found with the usage of $\alpha_y$ and $\beta_y$ of circumscribed ellipsoid.

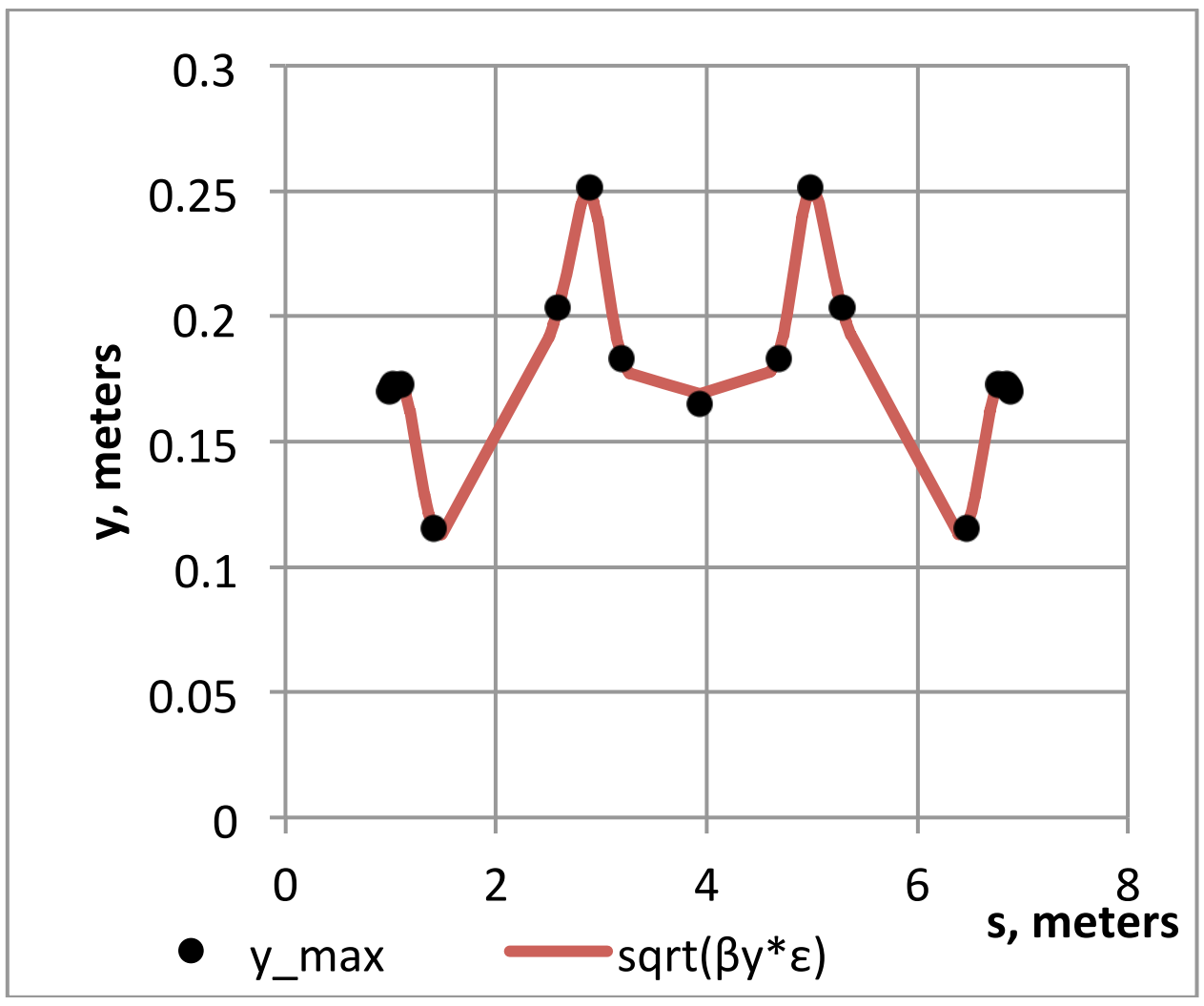

a)

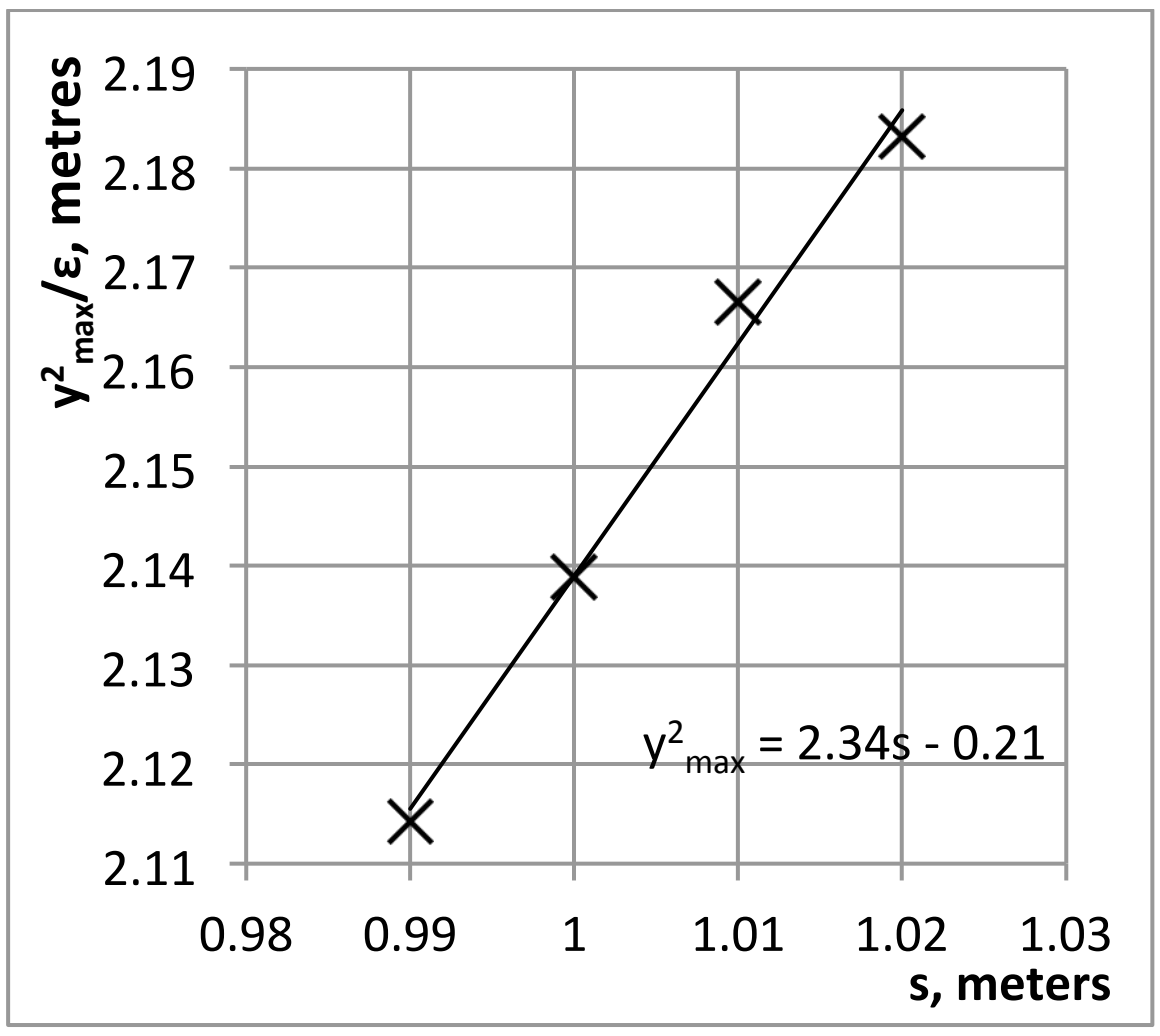

b)

Figure 11. Case with ellipse-like initial distribution: a) vertical beam size computed by tracking and square root of beta-function throughout linear region; b) value of $y^2_{max}/\varepsilon$ at the beginning of linear region. Slope of the plot is proportional to alpha-function ($\alpha = -\beta'/2$).

A similar comparison can be done for the case with non ellipse-like initial distribution (Fig. 10b). Figure 11a presents the vertical beam size and square root of $\beta_y$ of circumscribed ellipsoid. As one can see, the agreement between these values is worse than in the case with ellipse-like initial distribution but still satisfactory. Also in this case $(y^2_{max}/\varepsilon)' = 4.04$ and $\beta_y' = 4.12$.

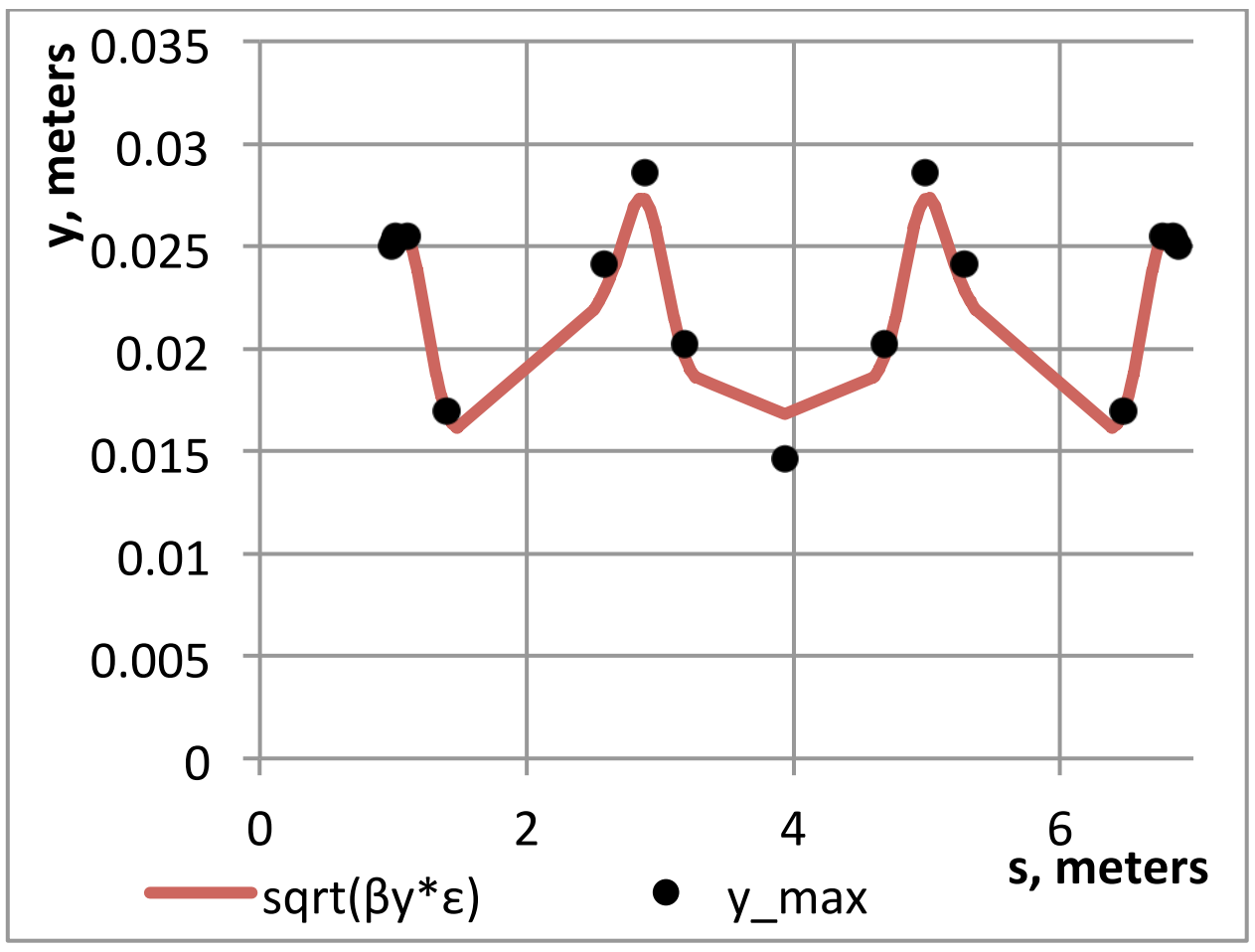

a)

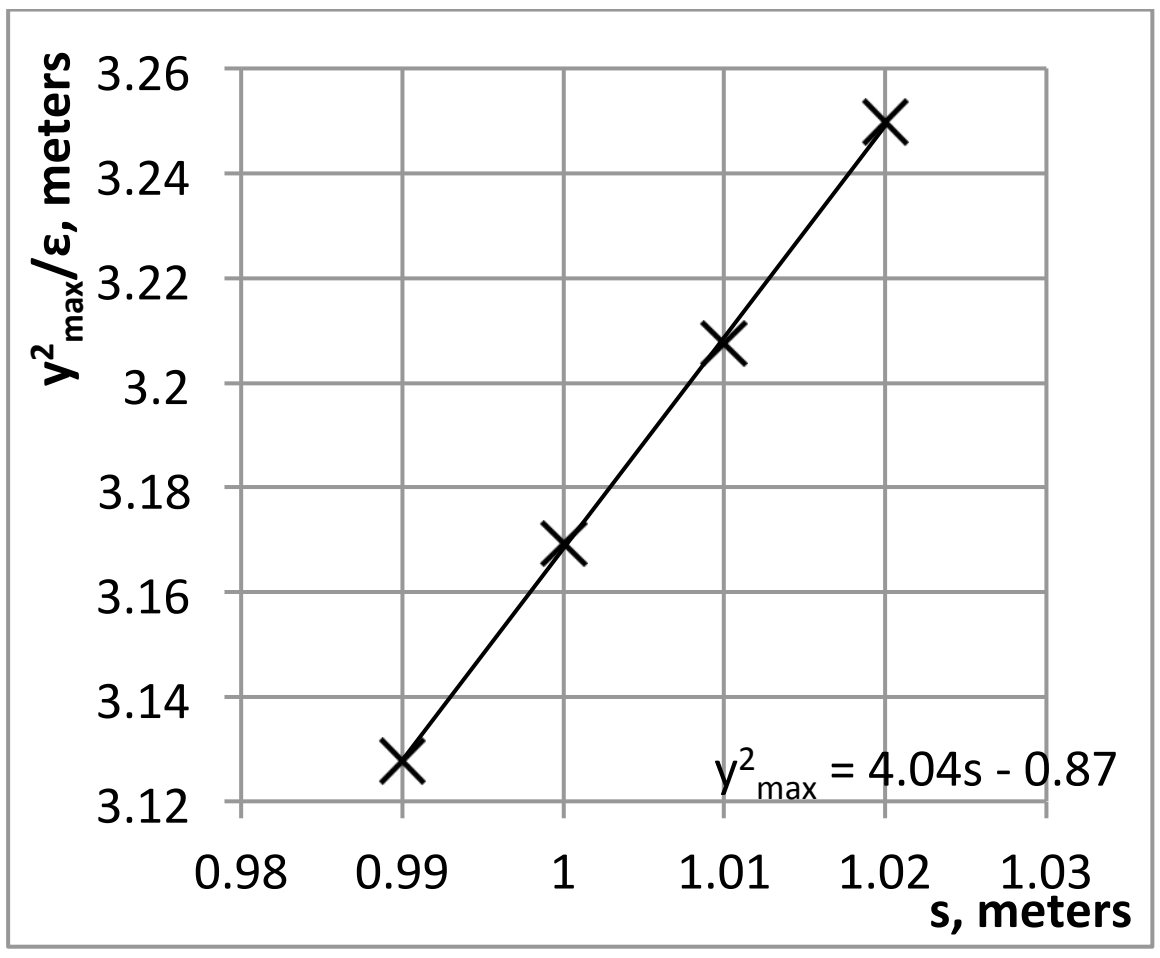

b)

Figure 12. Case with non ellipse-like initial distribution: a) vertical beam size computed by tracking and square root of beta-function throughout linear region; b) value of $y^2_{max}/\varepsilon$ at the beginning of linear region. Slope of the plot is proportional to alpha-function ($\alpha = -\beta'/2$).

Comparing ellipse-like and non ellipse-like cases one can mention that the better a circumscribed ellipse describes initial distribution, the better is proximity of finding the beam size by this method.

The same operations were carried out with 4-D case, when all of coordinates (x, px, y, py) are non-zero. Figure 13 presents the horizontal and vertical beam size and square root of $\beta_x$ and $\beta_y$ of circumscribed 4-D ellipsoid. One can see that the prediction of beam size given by $\beta_x$ and $\beta_y$ functions of circumscribed ellipsoid is accurate in this case. The difference between beam size and squared root of beta function is even less than in 2-D case with non-ellipse like shape of initial particle distribution (Fig. 12a).

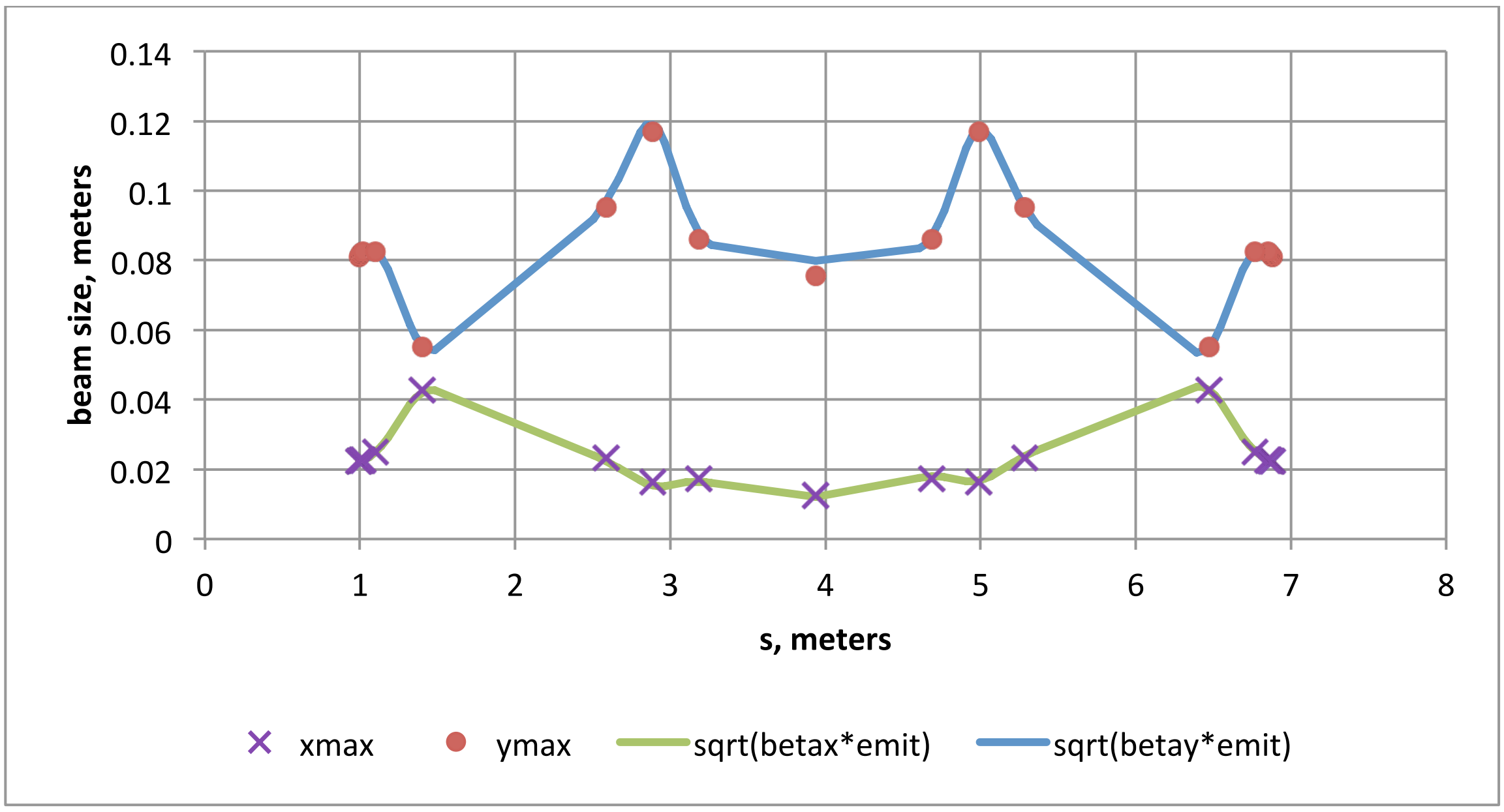

Figure 13. Maximum 'x' and 'y' coordinates of the beam computed by tracking and square roots of $\beta_x$ and $\beta_y$ functions throughout linear region.

Figure 14 presents the square beam size ('x' and 'y') divided by emittance as a function of longitudinal coordinate. For this case:

$$(x^2_{max}/\varepsilon_x)' = 2.83 \text{ and } \beta_x' = 2.91$$

$$(y^2_{max}/\varepsilon_y)' = 2.81 \text{ and } \beta_y' = 2.85$$

As one can see, the agreement of the 1st derivatives evaluated by tracking and using alpha-functions of circumscribed ellipsoid is also very good.

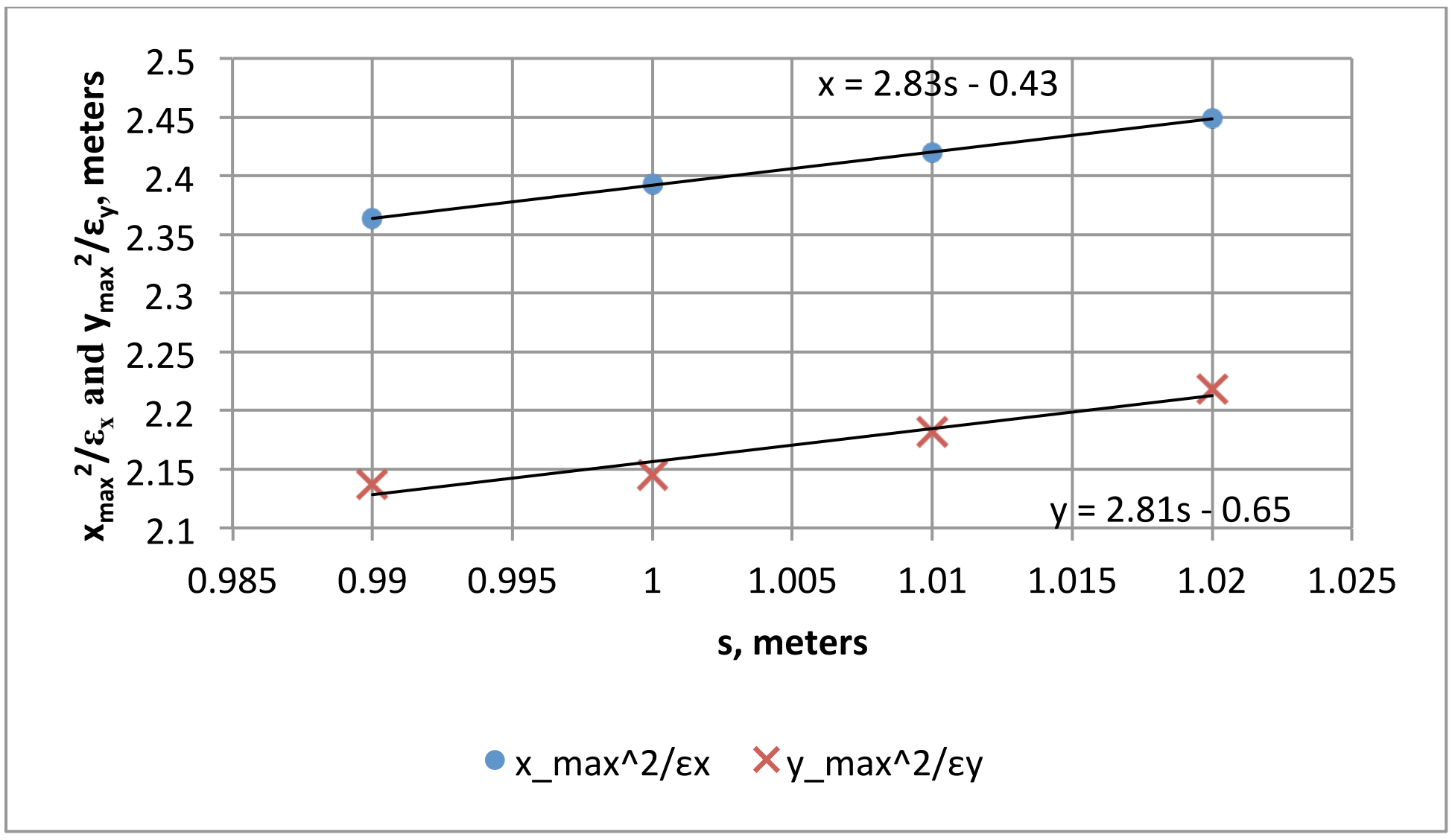

Figure 14. Value of $y^2_{max}/\varepsilon$ at the beginning of linear region. Slope of the plot is proportional to alpha-function ($\alpha_{x,y} = -\beta_{x,y}'/2$).

# Conclusions

1. The stable area of initial particle coordinates was analyzed as a function of non-linear lens strength.
    a. In a strong non-linear field, particles near the x-axis are typically lost at the physical aperture in quadrupoles;
    b. There is maximum strength of the non-linear lens for which particles survive.
2. There is no dynamic aperture limitation present in the system.
3. The found stable area is sufficient to observe an integer resonance crossing.
4. Possible tune spreads that could be achieved in the machine were analyzed. Maximum tune spreads are much larger than those for linear accelerators.
5. Chromaticity compensating sextupoles do not cause significant changes of the maximum attainable tune spread.
6. Method for estimating the beam size in linear regions of non-linear accelerator was proposed and tested for 2D and 4D cases. The precision of this method allows its use to predict the beam size for the machine lattice design studies.

# References

[1] 2S. Nagaitsev, A. Valishev, V. Danilov, "Nonlinear optics as a path to high-intensity circular machines", in Proceedings of 46th ICFA Advanced Beam Dynamics Workshop HB2010, 2010.

A. Valishev, V.S. Kashikhin, S. Nagaitsev, V.V. Danilov, "Ring for Test of Nonlinear Integrable Optics", in Proceedings of 2011 Particle Accelerator Conference, 2011

[2] MAD - Methodical Accelerator Design, http://mad.web.cern.ch/mad/